\documentclass[letterpaper]{article}

\usepackage{fontspec}
\usepackage{geometry}
\usepackage{setspace}
\usepackage[hidelinks]{hyperref}
\usepackage[style = chem-acs, doi=true]{biblatex}
\usepackage{graphicx}
\usepackage{float}
\newfloat{scheme}{htbp}{los}
\floatname{scheme}{Scheme}
\floatname{chart}{Chart}
\newfloat{graph}{htbp}{loh}

\usepackage{caption}
\usepackage{authblk}
\author[1]{Martin Anstett}
\author[1,2]{Ariyan Tavakoli}
\author[3]{Simone Mearini}
\author[3]{Yan Yan Grisan Qiu}
\author[3]{Vitaliy Feyer}
\author[2]{Peter Puschnig}
\author[1]{Martin Aeschlimann}
\author[4*]{Lu Lyu}
\author[1]{Hans Christian Schneider}
\author[4*]{Benjamin Stadtmüller}
\affil[1]{Department of Physics and Research Center OPTIMAS, RPTU University Kaiserslautern-Landau, Kaiserslautern, Germany}
\affil[2]{Institute of Physics, NAWI Graz, University of Graz, Graz, Austria}
\affil[3]{Peter Grünberg Institute (PGI-6), Jülich Research Centre, 52428 Jülich, Germany}
\affil[4]{Experimentalphysik II, Institute of Physics, University of Augsburg, Augsburg, Germany}

\title{Role of site-specific coordination for the structural and electronic properties of a Co-HATCN network on silver}
\date{*Email: lu.lyu@uni-a.de, benjamin.stadtmueller@uni-a.de}

\begin{document}

\maketitle

\begin{abstract}
Two-dimensional metal-organic porous networks (2DMOPNs) on metal surfaces provide a versatile platform for exploring emergent electronic and magnetic phenomena in designed surface nanostructures.
While the porous architecture is governed by coordination between molecular building blocks and metal centers, most 2DMOPNs utilize only one type of coordination site, leaving the impact of multiple coordination sites largely unexplored.

We address this question by investigating the structural formation of a Co-coordinated 2DMOPN on Ag(111) with the linker molecule hexaazatriphenylene-hexacarbonitrile (HATCN), which has two chemically distinct nitrogen coordination sites at the peripheral nitrile groups and pyrazine-type core nitrogen atoms. 
Scanning tunneling microscopy (STM) reveals the transformation from a simple honeycomb HATCN structure to a Co-coordinated porous network with three structurally and electronically inequivalent molecules in a 6:3:1 ratio.
The existence of two distinct coordination sites leads to selective Co coordination geometries for all three molecules, accompanied by site-specific charge transfer from Co into the former lowest unoccupied molecular orbital (LUMO), as revealed by momentum-resolved photoemission and density functional theory. Our results demonstrate that site-selective Co coordination not only controls the overall architecture of the porous network structure but also strongly influences the local electronic properties through site-specific metal-molecule interaction, providing a route to tailor the site-specific electronic structure of 2DMOPNs at the molecular level.

\end{abstract}

\section*{Keywords}

2D metal-organic porous networks, site-selective metal coordinations, molecular orbital alignment, photoemission orbital tomography, scanning tunneling microscopy, density functional theory  

\section{Introduction}

Two-dimensional metal--organic porous networks (2DMOPNs) on noble metal surfaces have emerged as a highly flexible platform for engineering the electronic properties of surfaces on the nanoscale~\cite{Barth2005,Barth2007,PiqueroZulaica2022}. The key control parameter is the long-range ordered porous architecture. This architecture determines the periodic surface potential landscape and thus governs the strength of quantum confinement of the two-dimensional electron gas of Shockley surface states and image potential states within the nanopores, as well as the coupling between neighboring pores~\cite{LoboCheca2009,PiqueroZulaica2022,PiqueroZulaica2019PRL,Lyu2026}. For instance, the band dispersion of the quantum-confined surface states depends on the pore size~\cite{Klappenberger2011,PiqueroZulaica2019Nanoscale}, the barrier width between adjacent pores~\cite{PiqueroZulaica2017}, and the geometry and symmetry of the scattering potential landscape~\cite{Kepcija2015,PiqueroZulaica2022}. These structural parameters can even be tuned by external stimuli such as temperature~\cite{Lyu2019}. Understanding the relationship between the structural motif and intralayer interactions within 2DMOPN structures is therefore crucial for achieving active control of the interfacial electron structure.

The most descriptive intralayer interaction for the overall network structure is the metal--ligand coordination, i.e., the interaction between the reactive molecular site and the linker atoms. These interactions are highly selective and directional, with interaction energies that can exceed those of mere van der Waals or hydrogen bonds~\cite{Barth2007}. As a result, metal--ligand coordination dominates the relative arrangement between the metallic linker atoms and the molecular building blocks, thus determining the overall periodicity and symmetry of the network structure. In addition, metal--ligand bonding substantially modifies the electronic structure of the organic building blocks. For strong electron acceptors, metal coordination induces a charge transfer into the lowest unoccupied molecular orbital (LUMO) of the ligand, whose magnitude depends on the type of metal center and its coordination number~\cite{Faraggi2012}. In extended 2D metal--organic networks, the hybridization between metal 3\textit{d} states and ligand $\pi$ orbitals even leads to the formation of hybrid dispersive bands~\cite{Baranowski2024,Mearini2024,Brandstetter2025}, and the charge transfer from the metal center to the ligands also alters the scattering potential of the metal centers for surface electrons~\cite{Klappenberger2011}.
So far, however, the coordination in most 2DMOPNs is mediated by only one type of coordination site of the organic ligand, for instance by carboxylate~\cite{Stepanow2004,Gambardella2009}, pyridyl~\cite{Vijayaraghavan2013,Umbach2014,PiqueroZulaica2019PRL}, or carbonitrile groups~\cite{Schlickum2007,Klappenberger2011,Abdurakhmanova2013}. In these systems, structural complexity arises mainly from a variation of the coordination number of the metal centers, as observed for the coexistence of three- and fourfold Co--carbonitrile nodes~\cite{Marschall2010} or for even more complex arrangements including quasicrystalline tilings~\cite{Ecija2013,Urgel2016}. In addition, chemically distinct coordination sites have so far been exploited for the selective binding of different metal species, for instance Gd atoms at the peripheral nitrile groups and Co atoms in the macrocycle of a porphyrin derivative~\cite{Urgel2015}. However, the competition among coordination sites of the same ligand for the same metal center, and its impact on the structural properties of the network, have remained largely unexplored.

Therefore, we investigate the structure formation of a cobalt-coordinated HATCN (hexaazatriphenylene-hexacarbonitrile) porous network structure on Ag(111). HATCN is a planar, highly symmetric $\pi$-conjugated molecule and a strong electron acceptor on noble metal surfaces~\cite{Broeker2010,Chakraborty2025}. On Ag(111), its adsorption is accompanied by a charge transfer from the substrate into the LUMO~\cite{Mueller2019}, which amounts to approximately one electron per molecule and lifts the twofold degeneracy of the gas-phase LUMO~\cite{Chakraborty2023}. Most importantly for our study, HATCN has two chemically distinct types of nitrogen atoms for metal coordination: six peripheral nitrile groups, which constitute the prototypical coordination motif of carbonitrile-based networks~\cite{Schlickum2007,Marschall2010} and six pyrazine-type nitrogen atoms of the hexaazatriphenylene core, which coordinate Cu atoms in Cu--HAT frameworks on coinage metal surfaces~\cite{CuHAT2024}. HATCN therefore represents an ideal model system to explore how the competition between two coordination sites governs the structure formation and the electronic properties of a metal--organic network.

In our work, we reveal the impact of Co coordination on the structural and electronic properties of a (sub-) monolayer of HATCN on Ag(111). Using scanning tunneling microscopy and low-energy electron diffraction, we observe a transformation from a simple honeycomb HATCN structure to a complex porous network with three structurally and electronically inequivalent molecules upon Co coordination. This transition is accompanied by a distinct, site-specific charge transfer from the Co atoms into the former LUMO, as determined by combining momentum-resolved photoemission and density functional theory. Together, these findings show that competition among coordination sites cannot only result in complex structural motifs in 2DMOPNs but also lead to site-specific modifications of the local molecular states within the network structures.

\section{Results and Discussion}

\begin{figure}[t]
  \centering
  \includegraphics[width=0.8\linewidth]{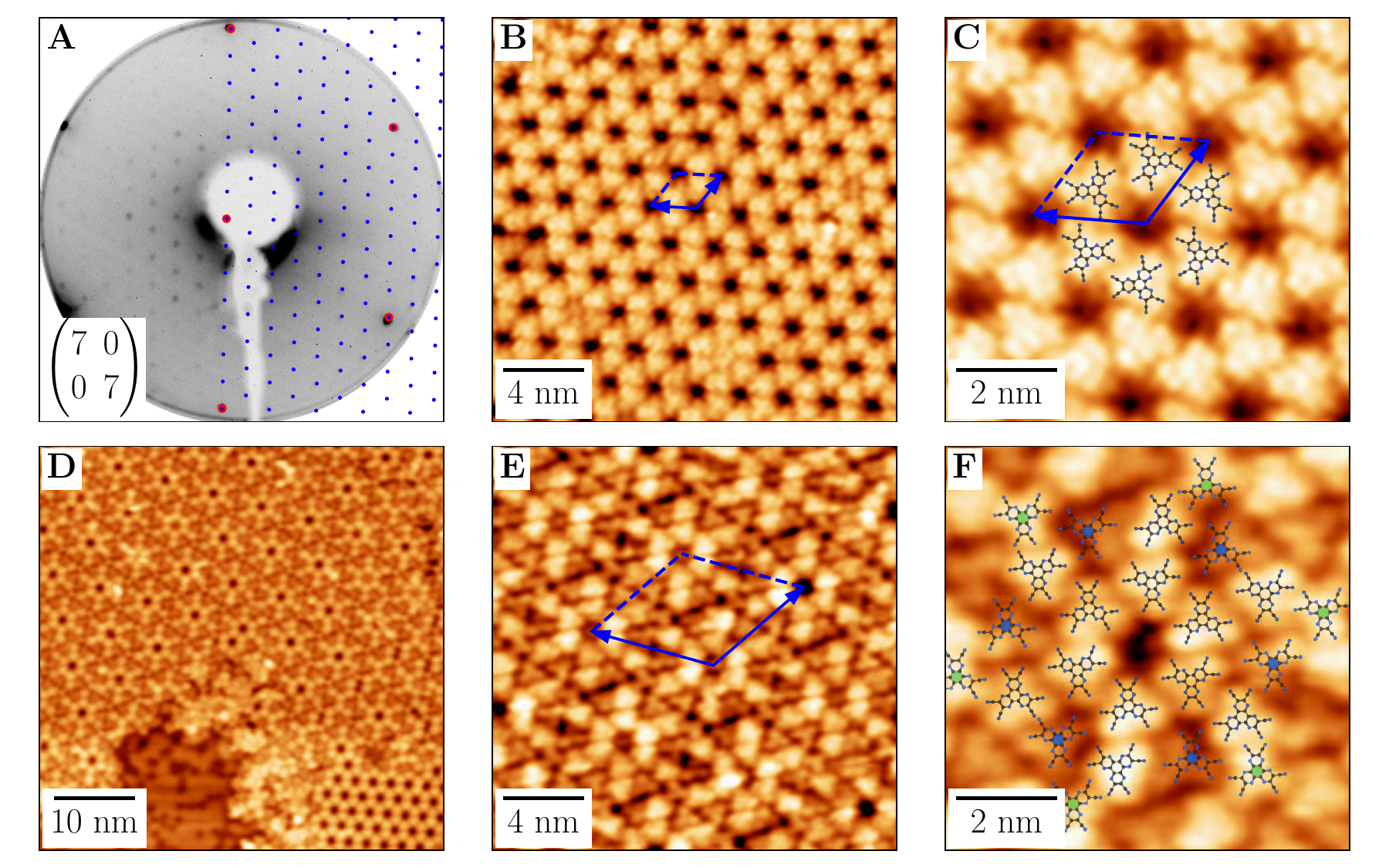}
  \caption{Structure of the HATCN and the Co-coordinated HATCN network on Ag(111). (A) LEED pattern of HATCN on Ag(111) recorded at 40\,eV, overlaid with the calculated diffraction spots of the $(7\times7)$ superstructure (blue) and the first-order Ag(111) spots (red). (B, C) STM images of HATCN on Ag(111) recorded at $T = 106$\,K, with the unit cell indicated in blue: (B) $V_\mathrm{t} = 1.2$\,V, $I = 50$\,pA; (C) $V_\mathrm{t} = -0.5$\,V, $I = 50$\,pA, with molecular models of six HATCN molecules superimposed. (D-F) STM images of the Co-HATCN network recorded at $T = 300$\,K. (D) Large-scale STM image ($V_\mathrm{t} = 1.5$\,V, $I = 30$\,pA) showing the coordination network next to a residual domain of the pristine honeycomb layer (lower right). (E) Close-up with the unit cell indicated in blue ($V_\mathrm{t} = 1.5$\,V, $I = 50$\,pA). (F) Magnified view with superimposed molecular models, in which the three molecular types are marked as uncolored, blue and green.
}
\label{fgr:STM_structure}
\end{figure}

We first characterize the pristine HATCN monolayer on Ag(111), which serves as the reference for the Co-coordinated network. The low-energy electron diffraction (LEED) pattern in \autoref{fgr:STM_structure}A exhibits a set of sharp superstructure spots, all of which are reproduced by a commensurate ($7\times7$) superlattice (blue spots) with respect to the Ag(111) atomic lattice (red spots), indicating a single long-range ordered phase. STM images in \autoref{fgr:STM_structure}B,C reveal a honeycomb arrangement of HATCN molecules. The unit cell (blue arrows) is a hexagonal lattice along the Ag(111) high-symmetry directions $\langle011\rangle$. The lattice length of 2.02\,nm corresponds to seven times the Ag surface lattice constant, in quantitative agreement with the LEED structure. The high-resolution image in \autoref{fgr:STM_structure}C resolves the threefold-symmetric shape of the individual molecules and shows that each unit cell contains two HATCN molecules rotated by $180^\circ$ with respect to each other, as indicated by the superimposed molecular models. The honeycomb motif agrees with previous reports of HATCN on Ag(111)~\cite{Mueller2019, Chakraborty2023}.

Deposition of Co atoms onto this monolayer and subsequent annealing at 480\,K induce a profound structural transformation (\autoref{fgr:STM_structure}D-F). The large-scale STM image in \autoref{fgr:STM_structure}D shows that most of the surface is covered by a new ordered phase, characterized by a hexagonal lattice of porous structure surrounded by star-like molecular assemblies. A small residual domain of the pristine HATCN honeycomb structure persists in the lower right of the image, which allows a direct comparison of both phases within the same field of view. The close-up in \autoref{fgr:STM_structure}E reveals that the new phase forms the hexagonal superlattice of 6.1\,nm, approximately three times that of the pure HATCN structure. Within this large unit cell, the molecules differ markedly in the apparent contrast, pointing to the presence of inequivalent molecular sites. 

To identify these sites, we superimposed molecular models onto the high-resolution image in \autoref{fgr:STM_structure}F and classified the molecules according to their apparent contrast. Three molecular species can be distinguished. Each pore is surrounded by a ring of six molecules, which is in turn enclosed by six further molecules. These twelve molecules (uncolored) constitute the star-like motif, and their mutual arrangement closely resembles that of the pristine pure HATCN structure. Six molecules that appear darker (blue) are interspersed between the uncolored molecules around the inner ring, and a third type of molecule (green) occupies the space between adjacent stars. The unit cell of the porous network thus contains twenty molecules, twelve uncolored, six blue, and two green, corresponding to a ratio of 6:3:1. Furthermore, the porous network formation alters the relative orientation of neighboring molecules. All molecules surrounding a green molecule adopt the same orientation (\autoref{fgr:STM_structure}F), whereas adjacent molecules in the pure HATCN phase are rotated by $180^\circ$ (\autoref{fgr:STM_structure}C). This local reorientation indicates that the incorporation of Co modifies the intermolecular binding motif rather than merely inserting Co atoms into the pristine lattice.

\begin{figure}[t]
  \centering
  \includegraphics[width=0.8\linewidth]{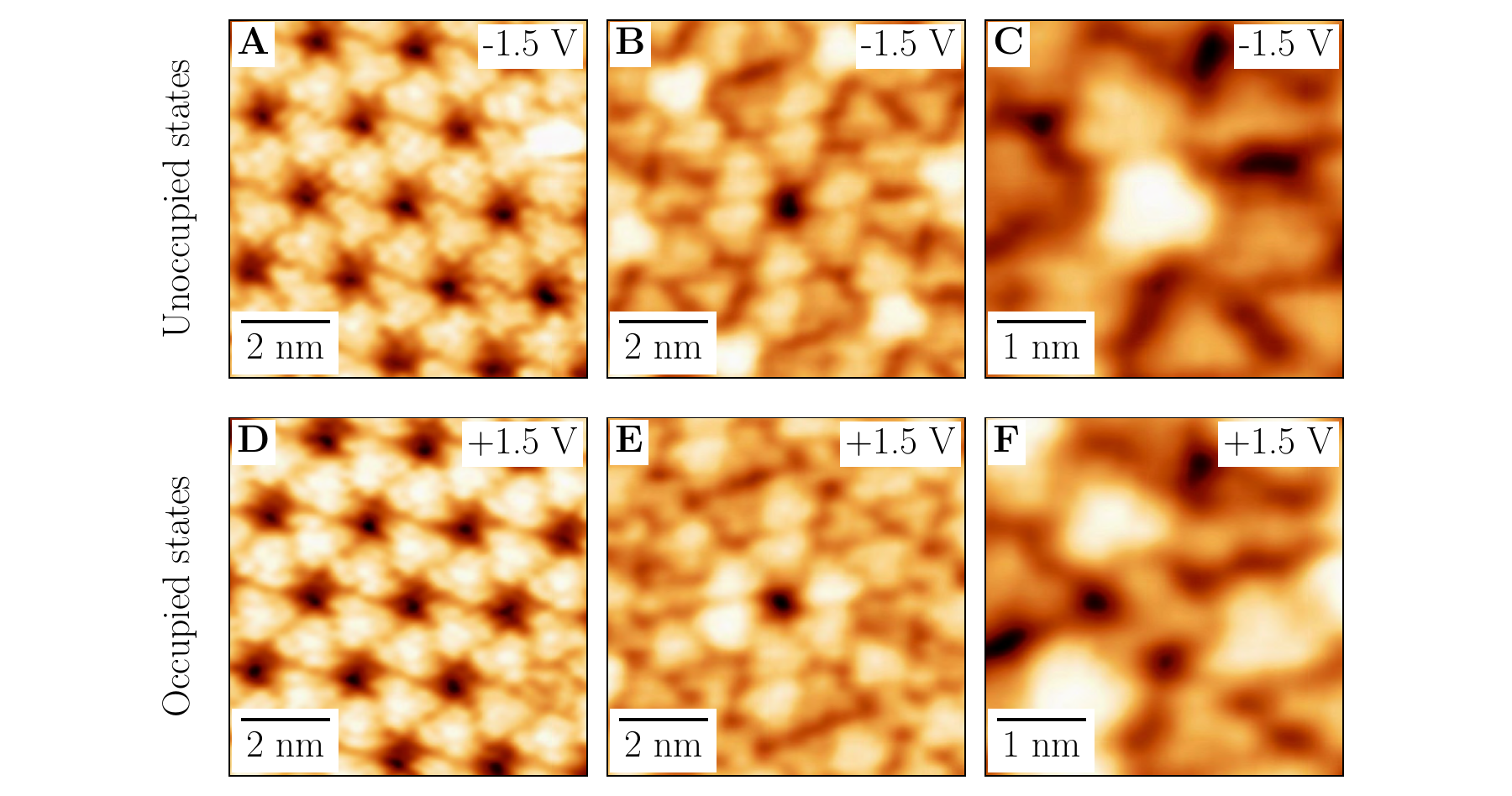}
  \caption{Bias-dependent STM contrast of (A, D) the HATCN layer and (B, C, E, F) the Co-HATCN network on Ag(111).
  Top row: $V_\mathrm{t} = -1.5$\,V (unoccupied states). Bottom row: $V_\mathrm{t} = +1.5$\,V (occupied states). The unit cell of the Co-HATCN network is shown in (B,E). (C,F) show magnified views centered on the green molecule. All images were recorded with $I = 50$\,pA.
  }
  \label{fgr:STM_bias_dependent}
\end{figure}

To examine whether the three species have different electronic states, we recorded STM images of both phases at opposite tip bias polarities (\autoref{fgr:STM_bias_dependent}). Since the bias voltage is applied to the tip, negative and positive tip biases probe the unoccupied and the occupied states of the sample, respectively. For the pure HATCN layer, the honeycomb contrast is essentially identical at $V_\mathrm{t}=-1.5$ and $+1.5$\,V, and all molecules appear equivalent (\autoref{fgr:STM_bias_dependent}A,D), in line with the two symmetry-related molecules of the unit cell. For the Co-HATCN network, by contrast, the appearance changes markedly with bias polarity (\autoref{fgr:STM_bias_dependent}B,C,E,F). At $V_\mathrm{t}=-1.5$\,V, the green molecules appear at the brightest protrusions of the network (\autoref{fgr:STM_bias_dependent}B). This is most evident in the magnified view in \autoref{fgr:STM_bias_dependent}C, where the green molecule in the image center stands out as a bright triangular feature. At $V_\mathrm{t}=+1.5$\,V, the contrast of the green molecule is strongly reduced relative to its uncolored neighboring molecules (\autoref{fgr:STM_bias_dependent}F), and the star-like motif formed by the uncolored molecules dominates the pattern (\autoref{fgr:STM_bias_dependent}E). The blue molecules appear dark at both bias polarities. The bias-dependent contrast inversion between green and uncolored molecules indicates the different molecular density of states near the region of the Fermi level. 

\begin{figure}[t]
  \centering
  \includegraphics[width=0.8\linewidth]{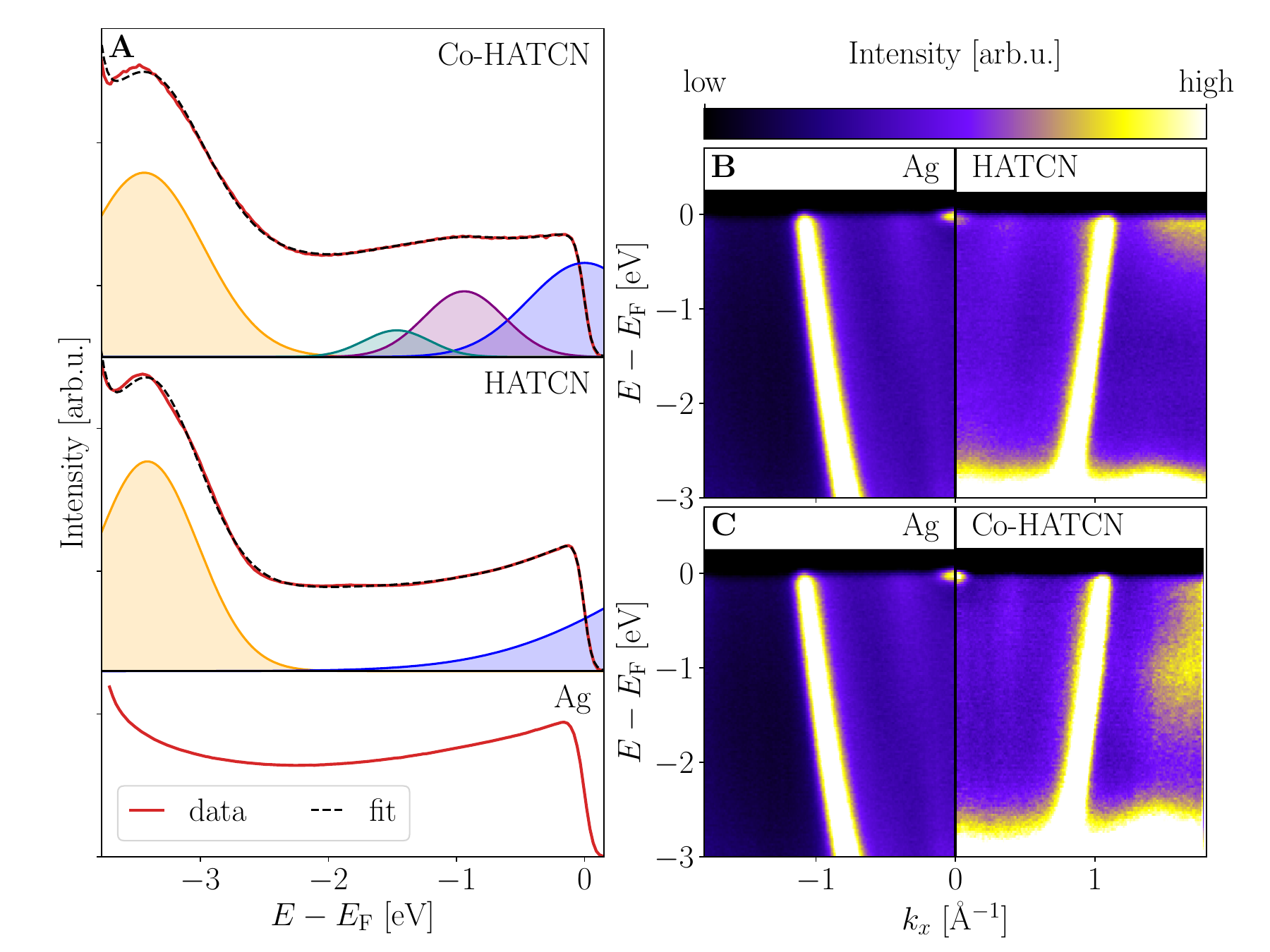}
  \caption{Photoemission of Ag(111), HATCN and Co-HATCN network on Ag(111) (h$\nu$ = 30\,eV, p-polarized light). (A) Momentum-integrated spectra (red) and fits (dashed black). Shaded areas show the fitted components: highest-occupied molecular orbital (HOMO, orange), partially filled former lowest-unoccupied molecular orbital (LUMO) at $E_\mathrm{F}$ (blue), and the two fully occupied former-LUMO components of the Co-HATCN network at $-0.94$\,eV (purple) and $-1.47$\,eV (green). For the Co-HATCN network, the three former-LUMO components are magnified by a factor of four. (B, C) Energy-momentum cuts along the $\bar{\Gamma}$--$\bar{K}$ direction of Ag(111) for bare Ag(111) (left halves) and for (B) pure HATCN and (C) Co-HATCN network (right halves).}
  \label{fgr:spectra}
\end{figure}

We now turn to the characterization of the electronic structure of the two phases. \autoref{fgr:spectra}A compares the momentum-integrated valence-band spectra of bare Ag(111), the pure HATCN monolayer, and the Co-HATCN network. The molecular spectra of HATCN and Co-HATCN exhibit two feature regions that are absent for bare Ag(111). The intense peak at $E- E_F=-3.4$\,eV stems from the HOMO of HATCN, which can be identified by comparing the observed and calculated orbital fingerprints in the momentum maps (\autoref{fgr:MM_kxky} and \autoref{fgr:momentum_map}). The second feature is located at the Fermi level and partially occupied below the Fermi edge. We assign it to the former LUMO, which is partially filled by charge transfer from the Ag(111) substrate~\cite{Chakraborty2023}. We quantified both contributions by fitting the spectra with Gaussian components multiplied by a Fermi-Dirac distribution, on top of an exponential background. Because only the occupied tail of the former LUMO is accessible in photoemission, its fitted peak position is at $E- E_F=0.4$\,eV.

In the Co-HATCN network, the spectrum changes qualitatively. While the HOMO peak retains the lineshape as in the pure HATCN case, the spectral peaks located between $-2$\,eV and $E_\mathrm{F}$ broaden and partially shift toward lower energies that cannot be described by a single component at the Fermi level (\autoref{fgr:spectra}A, top). Considering the three molecular species identified by STM (\autoref{fgr:STM_structure}F), we decomposed this region into three former LUMO components. Their integrated intensities were fixed to the 6:3:1 ratio of uncolored, blue and green molecules, assuming a comparable photoemission cross section per molecule within a unit cell. The fit yields one component at the Fermi level (blue), which closely resembles the partially filled former LUMO of the pure HATCN layer, and two components centered at $-0.94$\,eV (purple) and $-1.47$\,eV (green). These two components correspond to fully occupied LUMO states. The photoemission data thus reveal three distinct LUMO states within one network, i.e., a partially filled species as in pure HATCN, and two fully filled species at different energies. The fit of the integrated spectrum is based on the observed molecular species, and the momentum-resolved data discussed below provide signatures of these states in the momentum and energy space. 

The momentum-resolved data confirm that the additional states are of molecular features. \autoref{fgr:spectra}B,C show energy-momentum cuts along the $\bar{\Gamma}$--$\bar{K}$ direction of Ag(111), in which the left half of each panel displays bare Ag(111) and the right half corresponds to the respective molecular layer. For bare Ag(111), the prominent feature in this energy and momentum range is the steeply dispersing Ag \textit{sp} band at $|k_\parallel|\approx1$\,\AA$^{-1}$. At larger momenta, the substrate is free of states. Upon adsorption of HATCN, additional intensity appears beyond the \textit{sp} band (\autoref{fgr:spectra}B), in the momentum region ($|k_\parallel|\approx1.5$\,\AA$^{-1}$) where the LUMO of HATCN has the main lobe, as shown in \autoref{fgr:momentum_map}A. For the pure HATCN layer, this state is only visible in the vicinity of the Fermi energy, consistent with the partially filled former LUMO. For the Co-HATCN network, emission lobes in the same momentum range extend down to about $-2$\,eV (\autoref{fgr:spectra}C), mirroring the additional spectral peaks in the integrated spectrum. The additional occupied states of the Co-HATCN network are therefore derived from the former LUMO of HATCN. The more filled states indicate additional charges most likely provided by the Co atoms, pointing to a site-specific hybridization and charge transfer between Co and the former LUMO. 

The combined STM and photoemission results demonstrate three distinct environments of HATCN within the Co-HATCN network. STM resolves three molecular species with different bias-dependent contrast, and photoemission identifies three LUMO components near the Fermi level. Both observations indicate that the molecules differ in their coordinations to the Co atoms. However, the coordinating Co atoms could not be seen in our STM images at different bias voltages. This may be attributed to the spatial distribution of the Co states, which strongly overlap with the molecular orbitals, as well as to the limited spatial resolution of STM at room temperature. The possible Co positions are nevertheless constrained by the molecular structure of HATCN, which offers two chemically distinct types of nitrogen atoms for metal coordination, i.e., the peripheral nitrile groups and the pyrazine-type nitrogens of the hexaazatriphenylene core. It is therefore likely that the two fully occupied LUMO states arise from the hybridization of Co with these two distinct coordination sites, whereas the component at the Fermi level remains unchanged, originating from the uncoordinated molecules. To further clarify how the coordination site determines the electronic structure of the molecules, we performed density functional theory (DFT) calculations for the pure HATCN superstructure and a simplified freestanding Co-HATCN model comprising the three different coordination motifs.

\begin{figure}[t]
  \centering
  \includegraphics[width=0.8\linewidth]{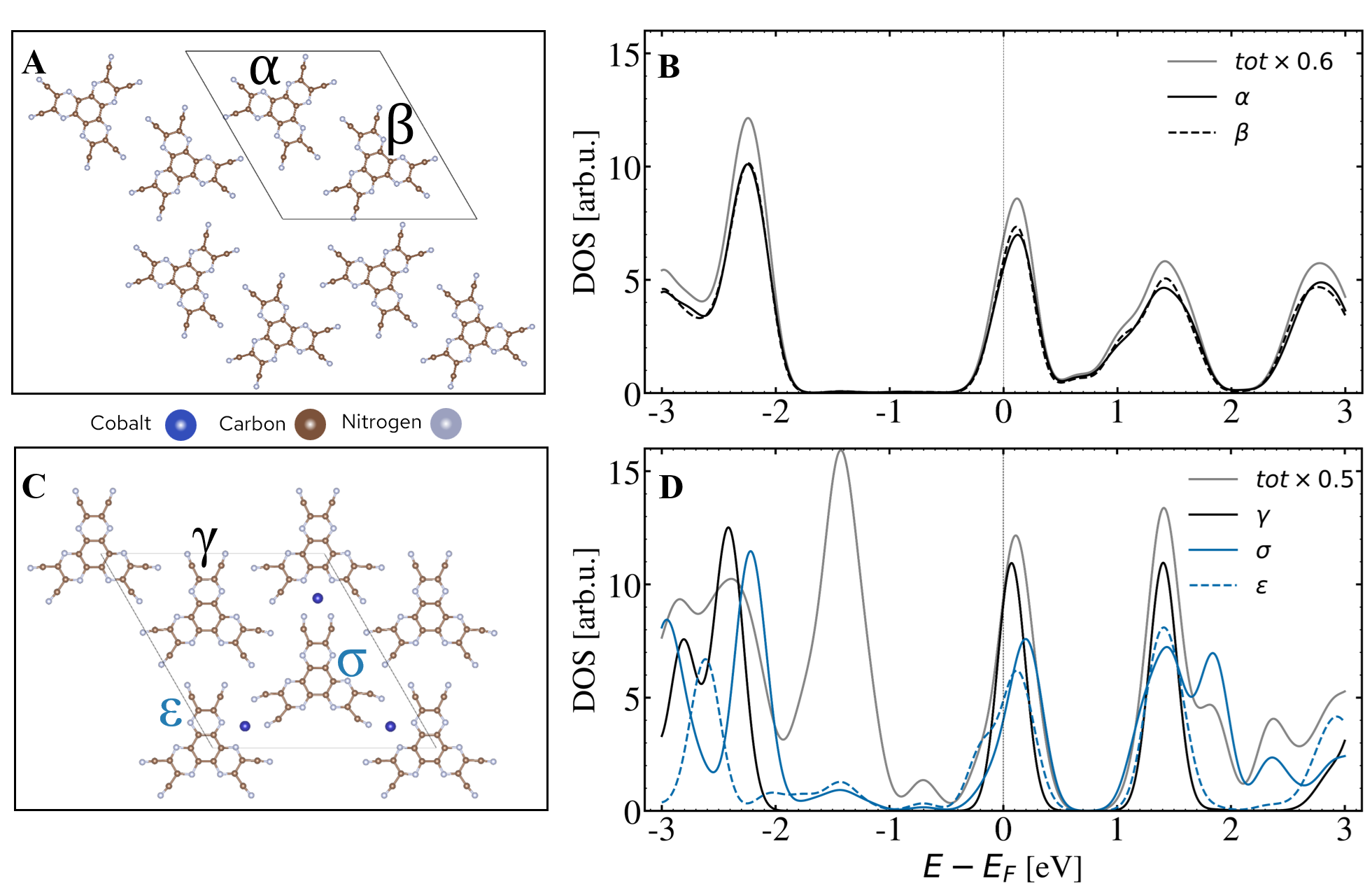}
  \caption{DFT models and molecule-projected LDOS. (A) Geometry optimized structure of the HATCN superstructur on Ag(111) (Ag atoms omitted for clarity), with the two molecules $\alpha$ and $\beta$ in the unit cell rotated by $180^\circ$ with respect to each other. (B)  Projected LDOS of  molecule $\alpha$ (solid), molecule $\beta$ (dashed) and total DOS (gray, scaled by 0.6). (C) Proposed model of relaxed freestanding Co-HATCN network, containing three inequivalent molecular coordination environments: an uncoordinated molecule ($\gamma$), a molecule coordinated to Co via its peripheral nitrile groups ($\sigma$), and a molecule coordinated to Co via the pyrazine-type nitrogen groups ($\varepsilon$). (D) Projected LDOS of molecule $\gamma$ (black solid), molecule $\sigma$ (blue solid), molecule $\varepsilon$ (blue dashed) and total DOS (gray, scaled by 0.5). The atomic structures are visualized using \textsc{VESTA} program~\cite{VESTA}.}
  \label{fgr:DOS_comparision}
\end{figure}

Based on the experimental observations, it is evident that in the pure structure, adjacent molecules are rotated by 180$^\circ$. The coordination with Co leads to a star-like structure in which only the central molecules of the star retain the arrangement of the pure structure. In contrast to the pure structure, at the corner of the star, adjacent molecules have the same orientation. This difference in orientation of the molecules is likely due to the presence of the Co dopants which influence the adsorption landscape.  

Two representative models for the pure HATCN and Co-coordinated HATCN structures are illustrated in \autoref{fgr:DOS_comparision}A and C.
\autoref{fgr:DOS_comparision}A shows the model structure of pure HATCN on Ag(111).
To support the experimental observations, periodic DFT calculations were performed for the pure HATCN adsorbed on Ag(111).
The unit cell in \autoref{fgr:DOS_comparision}A consists of two HATCN molecules which are rotated relative to each other by 180$^\circ$.
The inequivalent molecules are labeled as $\alpha$ and $\beta$. In the DFT calculation of the adsorbed molecules, the surface was modeled by a four-layer Ag(111) slab.
Details of the computational methodology, i.e., geometry optimization and electronic self-consistent field (SCF), can be found in the Methods section. \autoref{fgr:DOS_comparision}B shows the local density of states (LDOS) projected onto the individual  $\alpha$ and $\beta$ molecules.  It shows that the electronic structures of both adsorbed HATCN molecules with different relative orientations show only minor differences. In particular, they adsorb in a very similar manner.

The STM measurements reveal a well-ordered molecular layer in which all HATCN molecules adopt a uniform orientation in the presence of cobalt atoms. Since the experimental unit cell of the Co-HATCN network contains twenty molecules and would require a commensurate substrate cell of about a thousand Ag atoms in a four-layer slab, a DFT calculation of the complete network on Ag(111) is computationally prohibitive. We therefore constructed a freestanding model that captures the local coordination environments rather than the complete superstructure as shown in \autoref{fgr:DOS_comparision}C. The model is based on a hexagonal cell with a superstructure size of 2.1\,nm, derived from the average intermolecular distance in \autoref{fgr:STM_structure}F. The proposed structure includes three Co atoms and three inequivalent molecules: a molecule without a Co partner ($\gamma$), a molecule coordinated to Co through its peripheral nitrile groups ($\sigma$), and a molecule coordinated to Co at the pyrazine-type nitrogen atoms ($\varepsilon$). 
We use this structural model as a basis for a qualitative DFT analysis of the electronic structure associated with the different coordination environments. The computational details are provided in the Methods section. The calculated LDOS of the three inequivalent molecules establishes a direct connection between the coordination geometry and the experimentally observed STM contrast.

\autoref{fgr:DOS_comparision}D shows the LDOS for these different coordination motifs. The differences in coordination lead to distinct local electronic structures. The LDOS of molecule $\gamma$ closely resembles that of pristine HATCN, confirming its weak interaction with cobalt. In contrast, both $\sigma$ and $\varepsilon$ molecules exhibit pronounced modifications of their electronic structure due to metal coordination. A key fingerprint of this interaction is the emergence of new electronic states between approximately $-0.5$\,eV and $-1.5$\,eV below the Fermi level. In the isolated HATCN system in \autoref{fgr:DOS_comparision}B , this energy window is characterized by an absence of molecular states. Upon coordination with cobalt, however, new hybridized states appear in this region, reflecting strong metal-ligand coupling. This is consistent with photoemission spectroscopy measurements, which detect additional molecular features within the same energy range. Overall, the combined STM-derived structural model and DFT analysis demonstrate that cobalt atoms induce spatially varying electronic modifications depending on the local coordination geometry, leading to distinct electronic fingerprints for $\gamma$, $\sigma$, and $\varepsilon$ molecules. To assess the robustness of the model, we additionally tested a slightly larger supercell with a size of 2.4\,nm. Details of the computational methodology can be found in the Methods section.
The corresponding results are presented in \autoref{fgr:DOS_comparision_model2} in the Supporting Information.

\section{Conclusion}
Our study reveals a pronounced impact of the two chemically distinct coordination sites of HATCN on the structural and electronic properties of the Co-coordinated HATCN network on Ag(111). The incorporation of Co atoms transforms the pristine $(7\times7)$ honeycomb structure of HATCN into a complex, long-range ordered star-like network with a lattice constant of $6.1\,$nm, which contains three structurally and electronically inequivalent molecular sites in a ratio of 6:3:1. This structural transition is accompanied
by a site-specific occupation of the former LUMO, which we attribute to the local Co-HATCN interaction.

We relate both the site-specific Co-HATCN interaction and the resulting complex network structure to the competition between the different coordination sites of HATCN. This competition does not yield a single preferred coordination motif. Instead, nitrile-coordinated, pyrazine-coordinated, and uncoordinated molecules coexist within one unit cell. The complexity of the Co-HATCN network therefore arises from the
site-selective coordination of chemically identical ligands. At the same time, the local coordination motif determines the occupation of the molecular frontier orbitals molecule by molecule, so that a single network hosts HATCN molecules in three distinct electronic configurations.

More generally, our results identify the competition between coordination sites of a single molecular building block as an additional design parameter for 2DMOPNs. Because metal coordination and the associated charge redistribution modify the local scattering potential of the network for surface electrons, site-selective coordination offers a route to tailor with molecular precision not only the architecture of metal-organic networks but also the potential landscape that governs the quantum confinement of surface electrons.

\section{Methods}\label{sec:method}

\subsection{Sample Preparation}

All STM experiments were performed in an ultra-high vacuum (UHV) system with a base pressure below $10^{-9}$\,mbar.
The Ag(111) single crystal was cleaned by repeated cycles of argon ion sputtering and annealing at about 820\,K.  
HATCN molecules (Sigma-Aldrich, purity > 98\%) were thermally deposited from a Knudsen-type cell at 280\,\textdegree C onto the clean Ag(111) held at room temperature. Subsequently, Co atoms were deposited using an electron-beam evaporator (FOCUS EFM3, flux = 40 nA). The sample was then annealed at 480\,K for 60\,min. Samples for photoemission were prepared in-situ at the NanoESCA endstation following the same procedure.

\subsection{Scanning Tunneling Microscopy (STM)}

STM measurements were performed in a constant-current mode with a variable-temperature STM system (VT-AFM XA, Omicron GmbH) at 106\,K or 300\,K, as indicated, using an electrochemically etched W-tip. Tunneling currents were 30-50\,pA.
The bias voltage was applied to the tip, and positive and negative biases correspond to current tunneling at the sample's occupied and unoccupied states, respectively. WSxM software was used to process the data~\cite{WSXM}.

\subsection{Momentum-resolved Photoemission}

Photoemission experiments were carried out at the NanoESCA beamline of the Elettra synchrotron (Trieste, Italy) using a momentum microscope (Focus GmbH NanoESCA II), which collects photoelectrons emitted into the entire hemisphere above the sample surface~\cite{Wiemann2011}.
Scanning the kinetic energy of the photoelectrons yields a three-dimensional data set $I(E,k_x,k_y)$, composed of constant-energy momentum maps. Momentum-integrated spectra were obtained by integrating each map over the full momentum field, and the energy-momentum cuts were extracted along the $\bar{\Gamma}$--$\bar{K}$ direction of the Ag(111) surface Brillouin zone. All data were recorded with p-polarized light at a photon energy of 30\,eV, incident at $65^\circ$ with respect to the surface normal. The sample was measured at a temperature of 89\,K and a base pressure below $1\times10^{-10}$\,mbar. The energy and momentum resolutions were 100\,meV and $\pm0.05$\,\AA$^{-1}$, respectively. To minimize beam-induced damage of the organic layer, the sample was continuously rastered during data acquisition using a motorized stage.

\subsection{DFT Calculations}

\textbf{Gas-phase HATCN:} The geometry and orbitals of an isolated HATCN molecule were calculated with the \textsc{ORCA} program package~\cite{ORCA} using the hybrid PBE0 exchange--correlation functional~\cite{PBE0} and the def2-TZVPP basis set. Dispersion interactions were included using the Grimme-D3 correction with zero damping. Tight self-consistent field and geometry optimization convergence criteria were applied.
\\
\textbf{Pure HATCN on Ag(111):} Periodic DFT calculations were performed using the \textsc{VASP} code~\cite{VASP1,VASP2}. We used the Perdew--Burke--Ernzerhof (PBE) generalized gradient approximation (GGA)~\cite{PBE} together with projector augmented--wave (PAW) pseudopotentials. Dispersion interactions were included using the
Grimme-D3 correction~\cite{Grimme2010}. A plane-wave
energy cutoff of $400$\,eV was used. The Ag(111) substrate was modeled
using a slab containing four atomic layers. The in-plane lattice vectors define a surface supercell of approximately $20.24 \times 17.53$\,\AA,
while the lattice vector along the surface normal has a length of $37.08$\,\AA. Electronic occupations were treated using Gaussian smearing with a width
of $0.05$\,eV. The electronic energy convergence criterion was set to
$10^{-6}$\,eV. The structure was optimized using the conjugate--gradient
algorithm until the remaining forces on all atoms were smaller than
$0.01$\,eV/\AA. The Brillouin zone was sampled using a
$\Gamma$--centered $2 \times 2 \times 1$ grid. Dipole corrections were
applied along the surface normal to correct for artificial electrostatic
interactions between periodic slabs.
\\
\textbf{Freestanding Co--HATCN:} We used the same \textsc{VASP} setup,
PBE functional, and Grimme-D3 correction described above for HATCN/Ag(111).
The localized Co $3d$ orbitals were treated using the rotationally
invariant Dudarev formulation of DFT+$U$~\cite{Dudarev}, with an
effective Hubbard parameter of $U_{\mathrm{eff}}=4$\,eV. The network was modeled in a periodic supercell with dimensions of $21.0 \times 18.18 \times 30.0$\,\AA.
The reciprocal lattice is sampled using a $\Gamma$-point centred $5 \times 5 \times 1$ Monkhorst--Pack grid~\cite{MonkhorstPack}. The plane-wave energy cutoff, Gaussian smearing, structural optimization method, energy and force convergence criteria,
and dipole correction were the same as those used for HATCN/Ag(111).

\section*{Supporting Information}\label{sec:Supporting_information}

\begin{itemize}
    \item Figure S1. Geometric sensitivity of the freestanding network model: optimized structure
    and LDOS of the freestanding Co-HATCN network with a superstructure size of 2.4\,nm, calculated using DFT+$U$.

    \item Figure S2. Experimental constant-energy momentum maps of
    bare Ag(111), HATCN on Ag(111), and Co-HATCN network on Ag(111).

    \item Table S1. Calculated frontier orbitals of gas-phase HATCN.

    \item Figure S3. Simulated momentum maps of isolated HATCN, HATCN on Ag(111),
    and the freestanding Co-HATCN network with a superstructure size of 2.1\,nm.
\end{itemize}

\section*{Acknowledgements}
The authors thank the Deutsche Forschungsgemeinschaft (DFG, German Research Foundation) for financial support through TRR 173 -- 268565370 Spin+X (projects A08, B14 and MGK).
We acknowledge Elettra Sincrotrone Trieste for providing access to the NanoESCA beamline (proposal 20245486), and we thank Alexander Schmid, Ralf Hemm, and Jannis Lessmeister for their support during the beamtime.
All DFT calculations were performed on the high-performance computing cluster Elwetritsch at RPTU Kaiserslautern-Landau. A CPU grant for the project ``Spin+SMM" through the Alliance for High Performance Computing Rheinland-Pfalz (AHRP) is gratefully acknowledged. Last but not least, the authors thank Dominik Brandstetter and Siegfried Kaidisch for help with computational aspects of the theoretical photoemission methodology in the early stages of this work.

\printbibliography

\newpage

\appendix

\section{Supporting Information}\label{sec:Supplement}
This supplement provides additional experimental results and computational details that support the main text. We examine the influence of supercell size on the structural and electronic properties of the freestanding Co-HATCN network and present supplementary photoemission spectra and momentum maps. We also describe the molecular orbital calculations and photoemission simulations used to investigate how adsorption on Ag(111) and Co coordination affect the electronic structure of HATCN.

\setcounter{figure}{0}                
\begin{figure}[ht]
  \centering
  \includegraphics[width=0.8\linewidth]{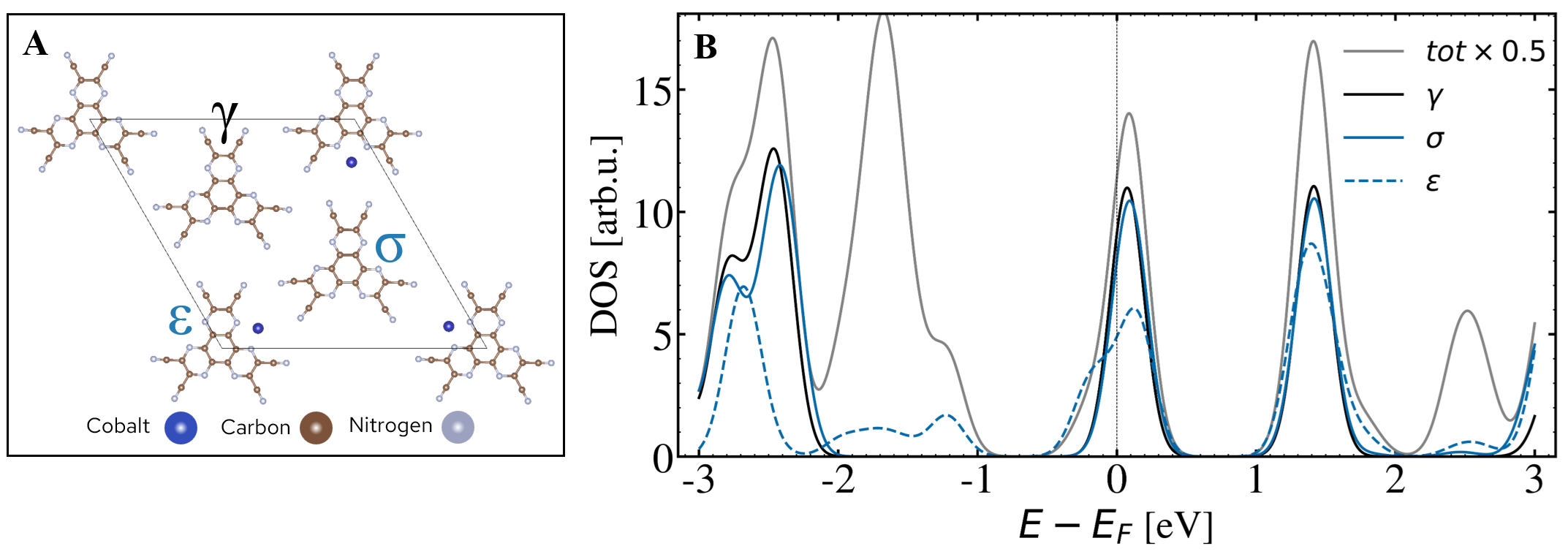}
  \renewcommand{\thefigure}{S1}
  \caption{(A) Optimized atomic structure of the freestanding Co-HATCN structure calculated using a larger supercell with a superstructure size of 2.4\,nm within the DFT+$U$ approach. (B) LDOS of the three HATCN molecules, labeled as $\gamma$, $\sigma$, $\varepsilon$. The gray solid line denotes the total DOS (gray, scaled by 0.5). The atomic structure is visualized using \textsc{VESTA} program~\cite{VESTA}.
}
  \label{fgr:DOS_comparision_model2}
\end{figure}
\autoref{fgr:DOS_comparision_model2} was calculated to assess the robustness of the model by considering a slightly larger supercell with a superstructure of size 2.4\,nm. Details of the computational methodology can be found in the Methods section.

\autoref{fgr:DOS_comparision_model2}A shows the optimized atomic structure of the 2.4\,nm supercell, while \autoref{fgr:DOS_comparision_model2}B presents the LDOS for the three HATCN molecules, labeled as $\gamma$, $\sigma$, and $\varepsilon$.

A key fingerprint of the Co-HATCN interaction is the emergence of new electronic states in the energy window between approximately $-1$\,eV and $-2$\,eV below the Fermi level. Within this energy range, only the molecule $\varepsilon$ exhibits a noticeable redistribution of the electronic states in its LDOS, indicating a stronger hybridization with the Co atoms. In contrast, the LDOS of the $\gamma$ and $\sigma$ molecules show no electronic states over the same energy interval. This similar behavior suggests that both molecules interact more weakly with the substrate and are therefore less strongly bonded to the Co atoms than the $\varepsilon$ molecule.

It is worth mentioning that, for the smaller lattice constant shown in \autoref{fgr:DOS_comparision}D, the electronic fingerprint of the Co atoms is clearly visible in the energy range between approximately $-0.5$\,eV and $-1$\,eV below the Fermi level. In contrast, for the slightly larger 2.4\,nm supercell in \autoref{fgr:DOS_comparision_model2}B, the electronic states of Co atoms shift to lower energies below the Fermi level. 
Furthermore, these cobalt-derived electronic states are found to hybridize predominantly with the $\varepsilon$ molecule, while no comparable hybridization is observed for the $\gamma$ and $\sigma$ molecules. 
The shift of the cobalt states to deeper energies in the larger supercell leads to a less favorable agreement with the experimentally observed spectral features. In contrast, the smaller supercell reproduces the experimentally observed energy range of the cobalt-induced states more accurately, indicating that the smaller unit cell provides a better description and is therefore in closer agreement with the experimental results.

\setcounter{figure}{1}
\begin{figure}[ht]
  \centering
  \includegraphics[width=0.8\linewidth]{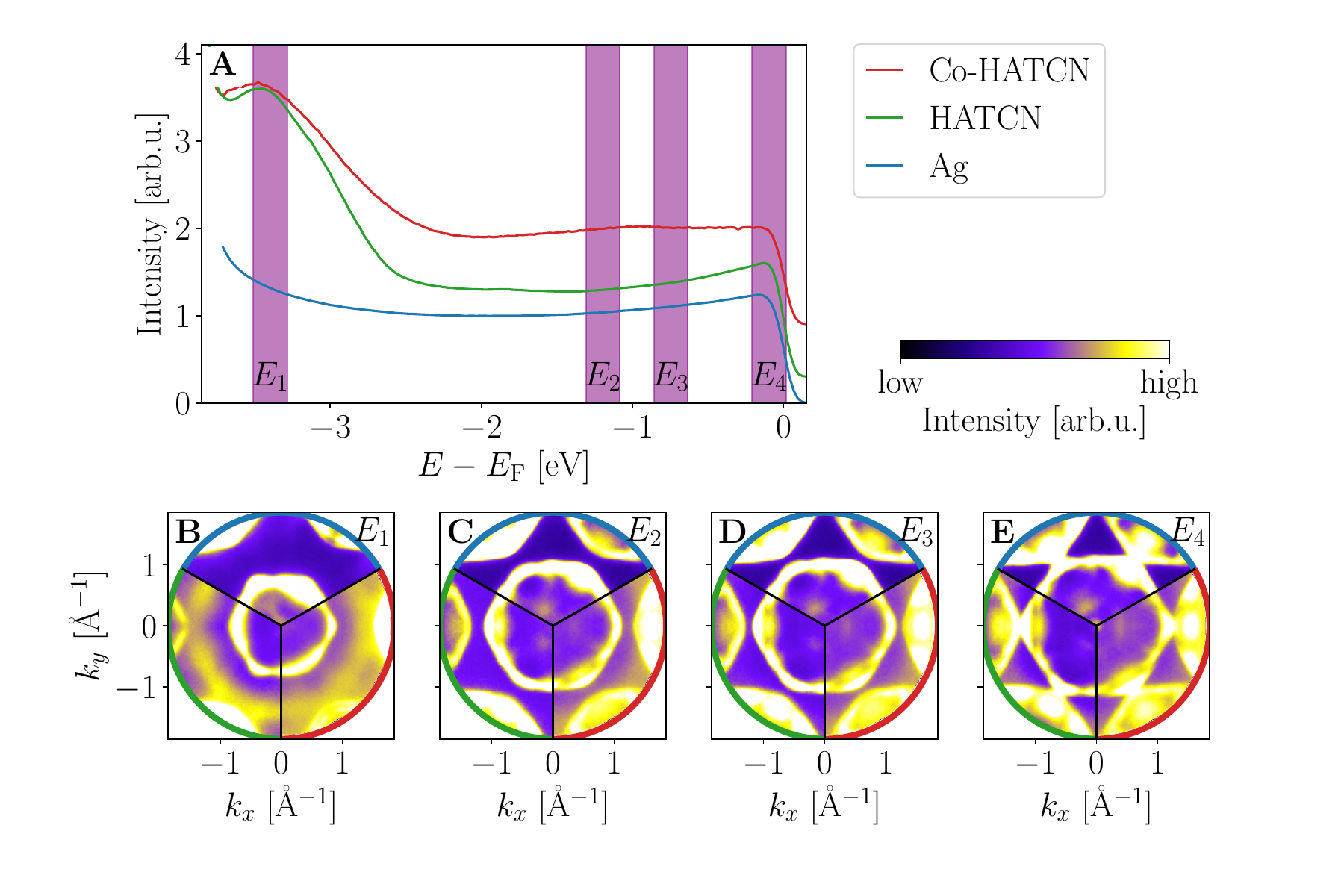}
  \renewcommand{\thefigure}{S2}
  \caption{(A) Momentum-integrated spectra of the Ag(111) (blue), HATCN on Ag(111) (green), and Co-HATCN on Ag(111) (red). Shaded bars mark the integration energy windows ($E_1$--$E_4$) for constant-energy maps. (B-E) Constant-energy maps at $E_1$--$E_4$, respectively, each composed of $120^\circ$ sectors from Ag(111) (blue frame), HATCN on Ag(111) (green frame), and the Co-HATCN network on Ag(111) (red frame). All measurements were performed at 89\,K with p-polarized light ($h\nu=30$\,eV).
  }
  \label{fgr:MM_kxky}
\end{figure}

\setcounter{figure}{2}
\begin{figure}[ht]
  \centering
  \includegraphics[width=0.7\linewidth]{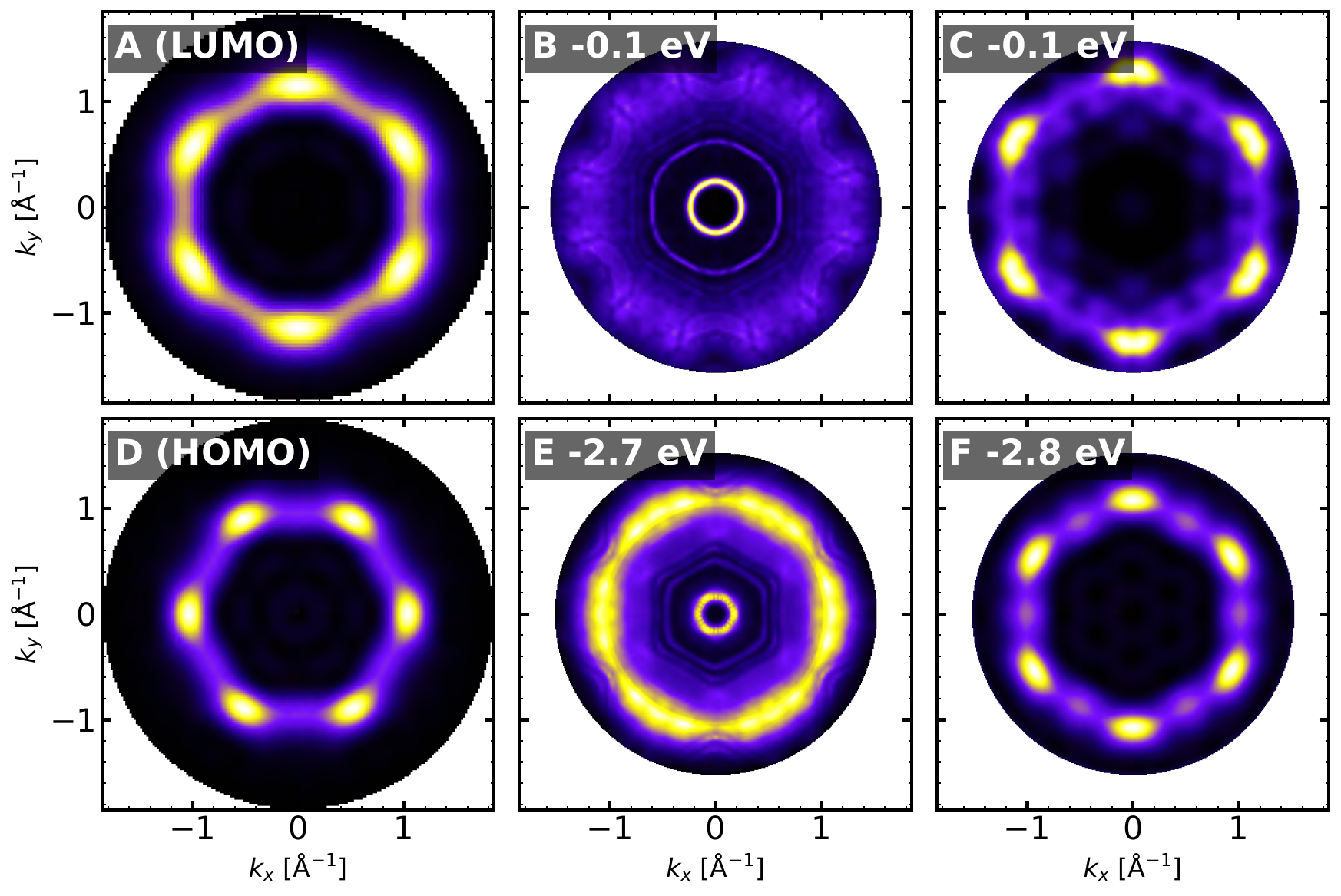}
   \renewcommand{\thefigure}{S3}
\caption{
Simulated photoemission momentum maps. (A, D) Isolated gas-phase HATCN molecule, obtained by combining the degenerate LUMO/LUMO$+1$ and HOMO/HOMO$-1$ orbitals, for the two symmetry equivalent molecules rotated by $180^\circ$. (B, E) HATCN adsorbed on Ag(111), calculated within the damped plane-wave final-state approximation for occupied state energy windows close to the Fermi level ($-0.1$\,eV) and at the maximum photoemission intensity ($\sim-2.70$\,eV). (C, F) The freestanding Co-HATCN network with a Co-HATCN superstructure of size 2.1\,nm, evaluated for occupied state energy windows at $-0.10$ and $\sim-2.80$\,eV below the Fermi level.
  } \label{fgr:momentum_map} 
\end{figure}
By combining DFT calculations with angle-resolved photoemission spectroscopy (ARPES)~\cite{Hufner2003} within the framework of photoemission orbital tomography (POT)~\cite{POT}, we exploit energy-resolved photoemission momentum maps to identify signatures of the molecular orbital that serve as fingerprints and can be used for orbital identification in ARPES experiments.
We first briefly review the underlying theory and assumptions.  

Here, we explain how we obtained the simulated momentum maps shown in \autoref{fgr:momentum_map}. 

\autoref{fgr:momentum_map}A and D show the momentum maps associated with the molecular orbitals of the lowest unoccupied molecular orbital (LUMO) and highest occupied orbital (HOMO) in an isolated pure HATCN system respectively. 

To simulate the momentum maps of the molecular orbitals, we performed DFT calculations for all three diffrent systems. Details of the computational methodology can be found in the Methods section. 
The frontier molecular orbital energies  of the isolated HATCN molecule obtained from the PBE0 calculations are summarized in \autoref{tab:hatcn_orbitals} which predict a doubly degenerate electronic structure for both the HOMO and LUMO of the isolated HATCN molecule, reflecting the high symmetry of the optimized molecular geometry. The calculated HOMO and LUMO energies are $-8.64$\,eV and $-5.58$\,eV, respectively, resulting in a gap of approximately 3.07\,eV. The negligible energy splitting between the degenerate states (i.e., below $\sim$ 1\,meV) originates from numerical precision.

\begin{table}[!htbp]
    \centering
    \caption[]{Frontier molecular orbital energies,  occupations and symmetry of the isolated HATCN molecule calculated using the PBE0 hybrid functional}
    \begin{tabular*}{\columnwidth}{@{\extracolsep{\fill}}c|ccc}
        \hline\hline
        Orbital & Symmetry &Occupation in e$^-$ & E [eV] \\
        \hline
        HOMO -1 & $e^{"+}$  & 2.0 & $-8.6483$ \\
        HOMO  & $e^{"+}$ & 2.0 & $-8.6481$ \\
        LUMO  & $e^{"+}$ & 0.0 & $-5.5807$ \\
        LUMO +1 & $e^{"+}$  & 0.0 & $-5.5803$ \\
        \hline\hline
    \end{tabular*}
    \label{tab:hatcn_orbitals}
\end{table}

To simulate the photoemission momentum maps, we employed the \texttt{kmap.py} program~\cite{kmap}. In these calculations, the gas-phase HATCN system consists of two symmetry-equivalent molecules, which are rotated by 180$^\circ$ with respect to each other. To properly account for this structural configuration, the momentum map shown in \autoref{fgr:momentum_map}A was constructed by summing the contributions of the LUMO and LUMO$+1$ orbitals of one HATCN molecule with those of the second molecule rotated by 180$^\circ$.

An analogous procedure was applied for the occupied states: the momentum map corresponding to the HOMO manifold shown in \autoref{fgr:momentum_map}D was obtained by combining the HOMO and HOMO$-1$ orbitals of one molecule with the symmetry-rotated counterparts of the second molecule. 

\autoref{fgr:momentum_map}B and E show the corresponding photoemission momentum maps of the pure HATCN molecule adsorbed on the Ag(111) surface. These calculations were performed for two different energy windows in the occupied region: one close to the Fermi level (i.e., $-0.10$\,eV) and a second one at approximately $-2.70$\,eV below the Fermi level, where the maximum photoemission intensity is observed.

By comparing these results with the HOMO and LUMO momentum maps of the isolated pure HATCN system  shown in \autoref{fgr:momentum_map}A and D, a clear molecular photoemission signature can be identified, demonstrating the persistence of characteristic molecular orbital features upon adsorption. In addition, distinct contributions originating from the Ag(111) substrate are clearly visible. 
We simulate the photoemission momentum maps in the POT formalism with an exponential damping ($\gamma$ = 0.5\,\AA$^{-1}$) factor to mimic the mean free path of escaping electrons~\cite{Chandrasekaran2020,Brandstetter2025}. Further details on this method can be found in~\cite{Lueftner2017}.

\autoref{fgr:momentum_map}C and F show the calculated photoemission momentum maps of the freestanding Co-HATCN system in a 2.1\,nm superstructure. The momentum maps were computed for two occupied state energy windows, one close to the Fermi level ($-0.10$\,eV) and the other at approximately $-2.80$\,eV below the Fermi level, corresponding to the maximum photoemission intensity in this energy region.

Comparison with the momentum maps of the isolated HATCN molecule shown in \autoref{fgr:momentum_map}A and D, as well as those of the HATCN/Ag(111) system in \autoref{fgr:momentum_map}B and E, reveals that the characteristic molecular photoemission signatures are largely preserved. At the same time, the spectral features are shifted by approximately 0.1\,eV toward higher binding energies. This energy shift provides clear evidence of electronic hybridization between the Co dopants and the HATCN molecules, while the overall momentum map symmetry indicates that the molecular orbital character remains largely intact.
\end{document}